\documentclass[conference, a4paper]{IEEEtran}

\renewcommand\IEEEkeywordsname{Index Terms}

\usepackage{graphicx}
\usepackage{caption}
\usepackage{cite}
\usepackage[hyphens]{url}
\usepackage[nolist]{acronym} 
\usepackage{pgfplots}

\usetikzlibrary{arrows,shapes,graphs,graphs.standard,quotes,arrows.meta,decorations.markings,positioning,calc}
\pgfplotsset{compat=newest}
\usepackage[bookmarks=false]{hyperref}
\usepackage{amsmath, amsbsy, amssymb}
\usepackage{amsthm}
\usepackage{multirow}
\usepackage{nicematrix}
\usepackage{booktabs}
\usepackage{nicefrac}

\usepackage{etoolbox} %

\usepackage[linesnumbered,vlined, ruled]{algorithm2e}

\usepackage{etoolbox}
\makeatletter
\patchcmd\algocf@Vline{\vrule}{\vrule \kern-0.4pt}{}{}
\patchcmd\algocf@Vsline{\vrule}{\vrule \kern-0.4pt}{}{}
\makeatother

\usepackage{marginnote}
\tikzset{>=latex}
\SetAlCapNameFnt{\footnotesize}
\SetAlCapFnt{\footnotesize}

\definecolor{mittelblau}{RGB}{0, 126, 198}
\definecolor{violettblau}{cmyk}{0.9, 0.6, 0, 0}
\definecolor{rot}{RGB}{238, 28 35}
\definecolor{apfelgruen}{RGB}{140, 198, 62}
\definecolor{gelb}{RGB}{255, 229, 0}
\definecolor{orange}{RGB}{244, 111, 33}
\definecolor{pink}{RGB}{237, 0, 140}
\definecolor{lila}{RGB}{128, 10, 145}
\definecolor{hellgrau}{RGB}{224, 224, 224}
\definecolor{mittelgrau}{RGB}{128, 128, 128}
\definecolor{dunkelgrau}{RGB}{80,80,80}
\definecolor{anthrazit}{RGB}{19, 31, 31}
\definecolor{darkgreen}{RGB}{34,139,34}
\definecolor{aqua}{RGB}{0, 255, 255}

\definecolor{lightgray}{RGB}{211,211,211}

\definecolor{neuesgruen}{RGB}{61, 173, 65}
\definecolor{dunklereshellgrau}{RGB}{176, 176, 176}
\definecolor{neuesgelb}{RGB}{255,160,0}
\definecolor{neuescyan}{RGB}{69,185,224}

\definecolor{tollesgruen}{RGB}{0,217,171}
\definecolor{tollesmagenta}{RGB}{197,67,143}

\definecolor{tollesgelb}{RGB}{255,199,95}
\definecolor{tollesrot}{RGB}{255,111,145}

\tikzset{
       vnd/.style={
        shape=circle,
        fill=black,
        draw,
        inner sep=0pt,
        minimum size=0.2cm},
        cnd/.style={
        shape=rectangle,
        fill=white,
        draw,
        minimum width=0.05mm,
        minimum height = 0.05mm}, 
         vndR/.style={
        shape=circle,
        fill=red,
        draw,
        inner sep=0pt,
        minimum size=0.2cm},
        cndR/.style={
        shape=rectangle,
        fill=white,
        draw=red,
        minimum width=0.05mm,
        minimum height = 0.05mm}
}

\IEEEoverridecommandlockouts

\renewcommand{\vec}[1]{\mathbf{#1}}

\newcommand{\sv}{\vec{s}}

\newcommand{\zerov}{\vec{0}}

\newcommand{\FF}{\mathbb{F}}

\newtheorem{remark}{Remark}
\newtheorem{definition}{Definition}

\newtheorem{proposition}{Proposition}
\newtheorem{example}{Example}

\begin{document}

\begin{NoHyper}
\title{Nested Symmetric Polar Codes}

\author{\IEEEauthorblockN{Marvin Rübenacke, Andreas Zunker, Felix Krieg, and Stephan ten Brink\\}
	\IEEEauthorblockA{
		Institute of Telecommunications, Pfaffenwaldring 47, University of  Stuttgart, 70569 Stuttgart, Germany 
		\\\{ruebenacke, zunker, krieg, tenbrink\}@inue.uni-stuttgart.de\\
	}
		\thanks{This work is supported by the German Federal Ministry of Education and Research (BMBF) within the project Open6GHub (grant no. 16KISK019).}
  }

\makeatletter
\patchcmd{\@maketitle}  %
{\addvspace{0.5\baselineskip}\egroup}
{\addvspace{-1.8\baselineskip}\egroup}
{}
{}
\makeatother

\maketitle

\begin{acronym}
\acro{ML}{maximum likelihood}
\acro{BP}{belief propagation}
\acro{BPL}{belief propagation list}
\acro{LDPC}{low-density parity-check}
\acro{BER}{bit error rate}
\acro{SNR}{signal-to-noise-ratio}
\acro{BPSK}{binary phase shift keying}
\acro{AWGN}{additive white Gaussian noise}
\acro{LLR}{Log-likelihood ratio}
\acro{MAP}{maximum a posteriori}
\acro{FER}{frame error rate}
\acro{BLER}{block error rate}
\acro{SCL}{successive cancellation list}
\acro{SCAL}{successive cancellation automorphism list}
\acro{SC}{successive cancellation}
\acro{BI-DMC}{Binary Input Discrete Memoryless Channel}
\acro{CRC}{cyclic redundancy check}
\acro{CA-SCL}{CRC-aided successive cancellation list}
\acro{PAC}{polarization-adjusted convolutional}
\acro{BEC}{Binary Erasure Channel}
\acro{BSC}{Binary Symmetric Channel}
\acro{BCH}{Bose-Chaudhuri-Hocquenghem}
\acro{RM}{Reed--Muller}
\acro{RS}{Reed-Solomon}
\acro{SISO}{soft-in/soft-out}
\acro{3GPP}{3rd Generation Partnership Project }
\acro{eMBB}{enhanced Mobile Broadband}
\acro{CN}{check node}
\acro{VN}{variable node}
\acro{GenAlg}{Genetic Algorithm}
\acro{CSI}{Channel State Information}
\acro{OSD}{ordered statistic decoding}
\acro{MWPC-BP}{minimum-weight parity-check BP}
\acro{FFG}{Forney-style factor graph}
\acro{MBBP}{multiple-bases belief propagation}
\acro{URLLC}{ultra-reliable low-latency communications}
\acro{mMTC}{massive machine-type communications}
\acro{DMC}{discrete memoryless channel}
\acro{SGD}{stochastic gradient descent}
\acro{QC}{quasi-cyclic}
\acro{5G}{fifth generation mobile telecommunication}
\acro{SCAN}{soft cancellation}
\acro{LSB}{least significant bit}
\acro{MSB}{most significant bit}
\acro{AED}{automorphism ensemble decoding}
\acro{AE-SC}{automorphism ensemble successive cancellation}
\acro{PPV}{Polyanskyi-Poor-Verd\'{u}}
\acro{RREF}{reduced row echelon form}
\acro{PL}{permutation linear}
\acro{LTA}{lower triangular affine}
\acro{BLTA}{block lower triangular affine}
\acro{PTPC}{pre-transformed polar codes}
\acro{UPO}{universal partial order}
\end{acronym}

\begin{abstract}
In this paper, we propose a data-driven algorithm to design rate- and length-flexible polar codes.
While the algorithm is very general, a particularly appealing use case is the design of codes for \ac{AED}, a promising decoding algorithm for \ac{URLLC} and \ac{mMTC} applications.
To this end, theoretic results on nesting of symmetric polar codes are derived, which give hope in finding a fully nested, rate-compatible sequence suitable for \ac{AED}.
Using the proposed algorithms, such a flexible polar code design for automorphism ensemble \ac{SC} decoding is constructed, outperforming existing code designs for \ac{AED} and also the 5G polar code under \ac{CRC}-aided \ac{SCL} decoding.

\end{abstract}
\acresetall

\section{Introduction}
Polar codes \cite{ArikanMain} have been commercialized in the \ac{5G} standard \cite{polar5G2018}.
Due to their good performance in the short block-length regime and the availability of low-complexity, fast decoding algorithms, it is likely that also 6G communication systems employ polar codes as one of their main error-correcting codes \cite{rowshan2024channelcoding6g, proceedings2024trends}.

A particularly promising decoding approach is \ac{AED} \cite{rm_automorphism_ensemble_decoding, pilletPolarCodesForAED}.
It exploits the rich symmetry group of polar codes \cite{polar_aed} to decode in a highly parallelizable fashion, enabling significantly reduced decoding latency and complexity compared to \ac{SCL} decoding \cite{Kestel2023URLLC}.
However, for useful symmetries to be available, tailored code design has to be applied.
While such codes individually exist for almost all practical parameters in the short to medium block-length regime \cite{pillet2023distribution}, for applicability in flexible wireless communication standards, a single unified description of a family of polar codes suitable for \ac{AED} is desirable \cite{bits2023unified}.
In particular, the polar code standardized in \ac{5G} is characterized by a nesting property \cite{Korada2010optimal}, that allows to derive codes with different lengths from the same sequence, making it flexible in both length and rate \cite{polarDesign5G}.
The design of such nested polar codes turns out to be a difficult problem and hence, previous work often resorted to machine-learning-based methods \cite{Huang2019NestedReinforcement, Li2021LearningNested, ankireddy2024nestedtransformer}.

To construct polar codes for \ac{AED}, previous work focused on the scenario where a desired symmetry group is predefined.
For example, \cite{pillet2022AEDdesign} proposes a method to iteratively refine a code design until the desired symmetries are obtained.
A more direct methodology is proposed in \cite{ShabunovPredefinedAutomorphisms}, which is extended to rate-compatible sequences in \cite{Geiselhart2023ratecompatible}.
However, the restriction to a predefined symmetry group limits the degrees of freedom in designing the code and almost certainly results in compromises in the error-correcting performance.
Moreover, a construction of a family of nested polar codes suitable for \ac{AED} has not been yet proposed.
Therefore, the goal of this paper is to construct a unified, nested sequence of rate- and length-flexible symmetric polar codes. Similarly to \cite{geiselhart2022graphsearch}, the algorithm is based on graph search on actual error-rate performance data. Our contributions are summarized as follows:
\begin{itemize}
    \item We derive theoretical results for the nesting property of symmetric polar codes.
    \item We propose a novel data-driven, general design algorithm for rate-compatible polar codes based on the shortest-path algorithm, which is then extended to nested polar codes.
    \item As an application, the algorithm is used to construct a nested polar code design for \ac{AED}, supporting lengths $32\le N\le 256$ and a wide range of code rates.
\end{itemize}

\section{Preliminaries}

\subsection{Notation}
Vectors and matrices are denoted by lowercase and uppercase boldface letters, respectively.
The $n$-bit \ac{LSB}-first binary expansion of an integer $i,\; 0\le i < 2^n$, is denoted by $\hat{\boldsymbol{i}} \in \mathbb{F}_2^n$.

\subsection{Polar Codes}
Polar codes are constructed from the polar transform defined by the $N\times N$ Hadamard matrix $\boldsymbol G_N = \left[\begin{smallmatrix} 1 & 0 \\ 1 & 1 \end{smallmatrix}\right]^{\otimes n}$ with $N=2^n$, where $(\cdot)^{\otimes n}$ denotes the $n$-th Kronecker power.
The polar transform converts identical channels into polarized synthetic channels, which are either highly reliable or unreliable.
Depending on the code design, $K$ reliable channels are selected to carry information and their indices form the \emph{information set} $\mathcal{I}$. The other, unreliable channels form the \emph{frozen set} $\mathcal{F}$.
The generator matrix $\boldsymbol{G}$ of the resulting polar code is composed exactly of those rows 
of $\boldsymbol G_N $ indexed by $\mathcal{I}$. 

\subsection{Partial Order}

The synthetic channels after the polar transform exhibit a \ac{UPO}, i.e., independent of the transmission channel, some synthetic channels are always more reliable than others \cite{bardet_polar_automorphism}.
If the synthetic channel indexed by $j$ is at least as reliable as the one indexed by $i$, we denote this by $i \preccurlyeq j$. 
The \ac{UPO} can be defined by the following two rules:
\begin{enumerate}
    \item \textit{Left swap}: If $(\hat{i}_l,\hat{i}_{l+1})=(1,0)$, and  $(\hat{j}_l,\hat{j}_{l+1})=(0,1)$ for $l < n$, while all other $\hat{j}_{k}=\hat{i}_{k}$, $k\not\in\{l,l+1\}$, then $i \preccurlyeq j$. 
    In other words, if $\hat{\boldsymbol{j}}$ equals $\hat{\boldsymbol{i}}$ with a single ``1" bit moved to a more significant position, then $i \preccurlyeq j$.
    \item \textit{Binary domination}: If $\hat{i}_l\le\hat{j}_l$ for all $l \leq n$, i.e, $\hat{\boldsymbol{j}}$ is the same as $\hat{\boldsymbol{i}}$, but may have additional ``1" bits, then $i \preccurlyeq j$. 
\end{enumerate}
The remaining relations are derived from transitivity.

We say that a polar code with information set $\mathcal{I}$ follows the \ac{UPO} if $i \in \mathcal{I} $ implies that all $j \succcurlyeq i$ are also in $\mathcal{I}$. In other words, for all information channels, all more reliable channels also carry information.
Polar codes following the \ac{UPO} can be compactly described by their minimum information set $\mathcal{I}_\mathrm{min}$ which is the smallest set from which $\mathcal{I}$ can be fully generated using the partial order relations \cite{polar_aed}.
In the following, all considered polar codes follow this \ac{UPO}.

\subsection{Automorphisms of Polar Codes}
\begin{definition}[Automorphism Group]\normalfont
The (permutation) automorphism group $\operatorname{Aut}(\mathcal{C})$ of a code $\mathcal{C}$ is the set of permutations of the codeword symbols that map every codeword onto another (not necessarily different) codeword. That means,
\begin{equation*}
    \operatorname{Aut}(\mathcal{C}) = \left\{ \pi \in S_N\;\middle|\;\pi(\boldsymbol{c}) \in \mathcal{C}\; \forall \boldsymbol{c} \in \mathcal{C} \right\},
\end{equation*}
where $\pi(\boldsymbol{c})$ is the permuted version of $\boldsymbol{c}$, $N$ is the block length and $S_N$ is the group of all permutations of $N$ elements.
\end{definition}

Except for degenerate cases, the automorphisms of polar codes are \textit{affine permutations}.
\begin{definition}[Affine Permutation]\normalfont
An affine permutation is a permutation of the codeword symbols which can be written as
\begin{equation}
    \pi: i\mapsto j: \quad \hat{\boldsymbol{j}} = \boldsymbol{A}\hat{\boldsymbol{i}} + \boldsymbol{b} \label{eq:affine_perm}
\end{equation}
where $\boldsymbol{A}\in \mathbb{F}_2^{n\times n}$ is an invertible matrix and $\boldsymbol{b}\in \mathbb{F}_2^n$. 
Furthermore, the affine automorphism group of a code is the subgroup of the automorphism group consisting of affine permutations.
\end{definition}

For a polar code, the structure of its affine automorphism group can be found using \textit{stabilizers}.
\begin{definition}[Stabilizer]\normalfont
The stabilizer $\operatorname{Stab}(\mathcal{S})$ of a set $\mathcal{S}\subseteq \{0,1,\dots,2^n-1\}$ is the set of permutations $\sigma \in S_n$ with
\begin{equation*}
    f_\sigma(i) \in \mathcal{S} \quad \forall i \in \mathcal{S},
\end{equation*}
where $f_\sigma(i)$ permutes the binary digits of $i$, i.e.,
\begin{equation*}
    f_\sigma: i \mapsto j: \quad \hat{\boldsymbol{j}} = \sigma(\hat{\boldsymbol{i}}). 
\end{equation*}
\end{definition}

If $\mathcal{I}$ is the information set of a polar code compliant with the \ac{UPO}, then $\operatorname{Stab}(\mathcal{I})$ exhibits a block structure, i.e., bits are only permuted within contiguous blocks with block sizes $\boldsymbol{s} = [s_0, ..., s_{m-1}] $, where $m$ is the number of blocks \cite{polar_aed}. 
Moreover, let $\operatorname{PL}(\boldsymbol{s})$ be the subgroup of affine permutations with $\boldsymbol{A}$ being a permutation matrix that does not have any non-zero entries above the block diagonal given by $\boldsymbol{s}$, and $\boldsymbol{b}=\zerov$.
Furthermore, the full affine automorphism group of a polar code is given by the \ac{BLTA} group, i.e., invertible affine mappings where $\boldsymbol{A}$ is restricted to have no non-zero elements above the block diagonal defined by $\boldsymbol{s}$ \cite{polar_aed,LiBLTA2021}.

\subsection{Successive Cancellation Decoding}
The first, and arguably the simplest decoding algorithm proposed with polar codes is \ac{SC} decoding, introduced in \cite{ArikanMain}.
This algorithm determines the most likely value for each information bit $u_i$ in a successive order, i.e.,
\begin{equation}
    \hat{u}_i = \arg \max_{u_i} \operatorname{Pr}\{u_i | \hat{u}_0 \dots \hat{u}_{i-1}, \boldsymbol{y}\},
\end{equation}
where $\boldsymbol{y}$ denotes the received vector. For frozen bits $i\in \mathcal{F}$, $\hat{u}_i = 0$.
\Ac{SC} decoding can be improved in error-rate performance by keeping a list of the $L$ most likely decoding paths, called \ac{SCL} decoding.
Typically, an outer \ac{CRC} code is utilized to both improve the weight enumerators of the polar code and to select the correct codeword from the list \cite{talvardyList}. This is referred to as \ac{CA-SCL}.

\subsection{Automorphism Ensemble Decoding}
An alternative decoding strategy to \ac{SCL} decoding is \ac{AED} \cite{rm_automorphism_ensemble_decoding}.
It uses an ensemble of $M$ identical and independent constituent decoders  (e.g., \ac{SC} decoders), each working on a permuted version of the received sequence $\boldsymbol{y}$, where the permutations are selected from the (permutation) automorphism group of the code. After un-permuting the respective codeword estimates, the most likely candidate is selected as the decoding result.
As the subdecoders are identical and independent, the complexity of \ac{AED} is significantly smaller than that of a comparable \ac{SCL} decoder \cite{Kestel2023URLLC}.

\section{Nested Polar Codes}
\begin{definition}[Low and High Nested Subcodes]\normalfont
Let $\mathcal{I}$ be the information set of a polar code with length $N$.
Its low and high nested subcodes are the $\nicefrac{N}{2}$ length polar codes specified by 
$\mathcal{I}^{\ell} = \left\{i \in \mathcal{I}: i<\nicefrac{N}{2}\right\}$ and $\mathcal{I}^{h} = \left\{i-\nicefrac{N}{2}: i  \in \mathcal{I}, i \ge \nicefrac{N}{2}\right\}$, 
respectively.
\end{definition}

For information sets $\mathcal{I}$ designed for \ac{SC} decoding at a given \ac{SNR} (in particular, via density evolution \cite{moriDE}), it is easy to see that $\mathcal{I}^{\ell}$ corresponds to a polar code designed for a worse \ac{SNR}, while $\mathcal{I}^{h}$ is designed for a better \ac{SNR}.

More generally, given a reliability sequence $Q$ of the synthetic channels, their respective subsequences 
$Q^{\ell} = [q_i \in Q : q_i < \nicefrac{N}{2}]$ and $Q^{h} = \left[q_i-\nicefrac{N}{2} : q_i \in Q,\,q_i \geq \nicefrac{N}{2} \right]$
are also valid reliability sequences for the shorter polar codes of length $\nicefrac{N}{2} $ \cite{Korada2010optimal}.
This is the \textit{nesting property}, that gives rise to elegant rate and length flexible polar codes which can be specified from a single reliability sequence
$Q_{N_{\mathrm{max}}}$ for the maximal block length $N_{\mathrm{max}}$ \cite{polarDesign5G}.
Due to the simpler implementation, lower nesting is used typically in practice.
To construct a polar code $\mathcal{P}(N=2^{n},K)$ from $Q_{N_{\mathrm{max}}}$, apply the following steps:
\begin{enumerate}
    \item Extract the subsequence $Q' = [q_i \in Q_{N_{\mathrm{max}}}: q_i < N]$
    \item Select the last $K$ channels as information set $\mathcal{I} = Q'_{(N-K):N}$
\end{enumerate}

In the following, we derive some useful properties regarding the affine automorphism groups of nested polar codes.

\begin{proposition}[Symmetric Subcodes]\label{prop:subcode}\normalfont
Let $\mathcal{P}(N,K)$ be a polar code with block profile $\boldsymbol{s} = [s_1,\dots,s_m ]$. Then its high and low nested subcodes 
$\mathcal{P}^\ell(\nicefrac{N}{2},K^\ell)$ and $\mathcal{P}^h(\nicefrac{N}{2},K^h)$
have at least block profile $\boldsymbol{s}' = [s_1,\dots,s_m-1 ]$ each (with $\boldsymbol{s}' = [s_1,\dots,s_{m-1}]$ if $s_m=1$).
\end{proposition}
\begin{IEEEproof}
Consider the subgroup of stage permutations (or bit-permutations) that fix the position $n-1$, corresponding to the block profile $\tilde{\boldsymbol{s}} = [s_1,\dots,s_m-1,1]$. By definition, no permutation in this subgroup permutes indices from one half of the code to the other. Furthermore, observe that for all $\pi\in \operatorname{PL}(\tilde{\boldsymbol{s}})$ and $0\le i < \nicefrac{N}{2}$, $\pi(i)-i = \pi(i+ \nicefrac{N}{2})-i+ \nicefrac{N}{2}$, i.e., the permutation acts identically on both halves. Therefore, the upper and lower subcodes are stabilized at least by $P_{\tilde{\boldsymbol{s}}}$ restricted to length $\nicefrac{N}{2}$, which is exactly $\boldsymbol{s}'$.
\end{IEEEproof}

\begin{proposition}[Symmetric Supercodes]\label{prop:supercode}\normalfont
Let $\mathcal{P}(N=2^n,K)$ be a polar code with minimum information set $\mathcal{I}_\mathrm{min}$ and block profile $\boldsymbol{s} = [s_1,\dots,s_m ]$. Then longer codes $\mathcal{P}'(N',K')$ with $N'=2^t N$ and information set $\mathcal{I}'$ constructed from $\mathcal{I}_\mathrm{min}$ have block profile $\boldsymbol{s}' = [s_1,\dots,s_m+t]$.

\end{proposition}
\begin{IEEEproof}
From the binary domination property, we know that $\forall i \in \mathcal{I}$, also $i+\tau\cdot N\in \mathcal{I}'$ with $1\le \tau < 2^t$. Hence, the stabilizing permutations of $\mathcal{I}'$ corresponding to the first $n$ positions are identical to those of $\mathcal{I}$.
Now partition $\mathcal{I}'$ into sets according to the prefix (corresponding to the least significant $n-s_m$ bits) and the Hamming weight of the $s_m$ most significant bits. Within each set, in the last block all possible bit patterns with same or higher Hamming weight are generated by the left swap rule. Therefore, the rightmost block is of size $s_m+t$.
\end{IEEEproof}
\begin{example}\normalfont
Let $\mathcal{P}(128,60)$ be the polar code constructed from $\mathcal{I}_\mathrm{min}=\{27\}$ with $\sv=[3,4]$. Its subcodes $\mathcal{P}^\ell(64,19)$ and $\mathcal{P}^h(64,41)$ both have $\sv'=[3,3]$.
The $t=2$ supercode $\mathcal{P}'(512,376)$ constructed from $\mathcal{I}_\mathrm{min}=\{27\}$ has $\sv'=[3,6]$.

\end{example}

These two theoretic properties give hope for finding nested sequences of codes with large automorphism groups.

\section{Finding Rate Compatible Sequences}
\subsection{General Methodology}
To find sequences of polar codes, we employ a graph search algorithm. Similarly to \cite{geiselhart2022graphsearch}, we first construct a directed graph, where each vertex corresponds to a unique information set $\mathcal{I}$. 
Two vertices $\mathcal{I}_i$ and $\mathcal{I}_j$ are connected by an edge, if and only if $\mathcal{I}_i \subset \mathcal{I}_j$ and $|\mathcal{I}_j| = |\mathcal{I}_i| + 1$.
A rate compatible sequence $\mathcal{S}$ of polar codes is a path through this graph from $K=0$ to $K=N$.
Note that for start and end, there exists only a single possible code, namely the trivial codes that contain only the zero-vector or all vectors, respectively.

For each code, let $\mu(\mathcal{C})$ denote a performance metric that, without loss of generality, is to be minimized\footnote{The code $\mathcal{C}_0$ with $K=0$ is merely included for consistent notation, and we set $\mu(\mathcal{C}_0)=0$.}. For example, $\mu(\mathcal{C})$ can be the required \ac{SNR} in dB for the desired decoding algorithm to achieve a target \ac{BLER} $\epsilon$. 

We define the following metric to assess the performance of a rate compatible sequence.
\begin{definition}[Optimal Sequence]\label{def:optimal_sequence}\normalfont The optimal sequence $\mathcal{S}$ of polar codes with respect to the metric $\mu(\mathcal{C})$ minimizes 
\begin{equation}\label{eq:pathmetric}
    \mu(\mathcal{S}) = \sum_{\mathcal{C} \in \mathcal{S}} \mu(\mathcal{C}).
\end{equation}
\end{definition}

\begin{proposition}[Optimality]\label{prop:shortest_path}\normalfont
The optimal sequence according to Definition \ref{def:optimal_sequence} is the shortest path through the graph of polar codes where each edge is associated with the cost
\begin{equation}\label{eq:edgecost}
    c(\mathcal{C}_i, \mathcal{C}_j) = \begin{cases} \frac{\mu(\mathcal{C}_i)+\mu(\mathcal{C}_j)}{2}&\text{if }\mathcal{C}_i\text{ and }\mathcal{C}_j\text{ are connected} \\ \infty & \text{else}\end{cases}.
\end{equation}
\end{proposition}
\begin{IEEEproof}
Observe that we can rewrite (\ref{eq:pathmetric}) as
\begin{align}
 \mu(\mathcal{S}) &= \sum_{\mathcal{C} \in \mathcal{S}} \mu(\mathcal{C})\nonumber\\
    &=\frac{1}{2}\mu(\mathcal{C}_0)+ \sum_{(\mathcal{C}_i,\mathcal{C}_j) \in \mathcal{S}} \frac{\mu(\mathcal{C}_i)+\mu(\mathcal{C}_j)}{2} +\frac{1}{2}\mu(\mathcal{C}_N),\nonumber
\end{align}
where $\mathcal{C}_0$ and $\mathcal{C}_N$ are the trivial codes contained in every sequence. Hence, the shortest path with the edge cost according to equation (\ref{eq:edgecost}) also minimizes (\ref{eq:pathmetric}).
\end{IEEEproof}
Note that the total cost of the path can be interpreted as the area under the path, when the vertices are arranged as $(x,y) = (\dim(\mathcal{C}), \mu(\mathcal{C}))$. This property is depicted in Fig.~\ref{fig:graph}.
Proposition \ref{prop:shortest_path} allows us to find a sequence of rate compatible codes efficiently and deterministically from a dataset using Dijkstra's algorithm \cite{Dijkstra1959}.
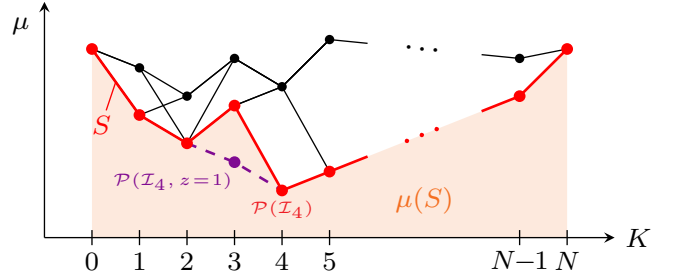
\begin{figure}[t]
    \centering
    \resizebox{\linewidth}{!}{\begin{tikzpicture}[>=stealth]

    \coordinate (D1) at (1.0, 1.8);
    \node[circle, fill=black, inner sep = 1pt] at (D1) {};
    \coordinate (D2) at (1.5, 1.5);
    \node[circle, fill=black, inner sep = 1pt] at (D2) {};
    \coordinate (D3) at (2.0, 1.9);
    \node[circle, fill=black, inner sep = 1pt] at (D3) {};
    \coordinate (D4) at (2.5, 1.6);
    \node[circle, fill=black, inner sep = 1pt] at (D4) {};
    \coordinate (D5) at (3.0, 2.1);
    \node[circle, fill=black, inner sep = 1pt] at (D5) {};
    \coordinate (DNmin1) at (5.0, 1.9);
    \node[circle, fill=black, inner sep = 1pt] at (DNmin1) {};
        
    \draw[black] (D5) -- ($(D5) + 0.2*(DNmin1) - 0.2*(D5)$);
    \draw[black] (DNmin1) -- ($(D5) + 0.8*(DNmin1) - 0.8*(D5)$);
        
    \path (D5) -- node[black, sloped]{\ldots} (DNmin1);
        
    \coordinate (start) at (0.5, 2.0);
    \coordinate (S1) at (1.0, 1.3);
    \coordinate (S2) at (1.5, 1.0);
    \coordinate (S3) at (2.0, 1.4);
    \coordinate (S3p) at (2.0, 0.8);
    \coordinate (S4) at (2.5, 0.5);
    \coordinate (S5) at (3.0, 0.7);
    \coordinate (SNmin1) at (5.0, 1.5);
    \coordinate (end) at (5.5, 2.0);
        
    \draw[black] (start) -- (D1) -- (S2) -- (D3) -- (D4) -- (D5);
    \draw[black] (start) -- (S1) -- (D2) -- (D3) -- (D4) -- (D5);
    \draw[black] (start) -- (S1) -- (S2) -- (S3) -- (D4) -- (D5);
    \draw[black] (start) -- (S1) -- (S2) -- (S3) -- (D4) -- (S5);
    \draw[black] (start) -- (D1) -- (D2) -- (D3) -- (D4) -- (D5);
    \draw[black] (DNmin1) -- (end);

    \begin{scope}
        \path[fill=orange, opacity=0.15] (0.5,0) -- (start) -- (S1) -- (S2) -- (S3) -- (S4) -- (S5) -- (SNmin1) -- (end) -- (5.5,0) -- cycle;
    \end{scope}
    
    \node[circle, fill=lila, inner sep = 1.25pt] at (S3p) {};
    \draw[lila, thick, dashed] (S2) -- (S3p) -- (S4);
    \node[lila] at(1.35,0.6) {\tiny $\mathcal{P}(\mathcal{I}_4,z\kern-.3em=\kern-.3em1)$};
    \node[rot] at(2.5,0.3) {\tiny $\mathcal{P}(\mathcal{I}_4)$};
        
    \node[circle, fill=rot, inner sep = 1.25pt] at (start) {};
    \node[circle, fill=rot, inner sep = 1.25pt] at (S1) {};
    \node[circle, fill=rot, inner sep = 1.25pt] at (S2) {};
    \node[circle, fill=rot, inner sep = 1.25pt] at (S3) {};
    \node[circle, fill=rot, inner sep = 1.25pt] at (S4) {};
    \node[circle, fill=rot, inner sep = 1.25pt] at (S5) {};
        
    \node[circle, fill=rot, inner sep = 1.25pt] at (SNmin1) {};
    \node[circle, fill=rot, inner sep = 1.25pt] at (end) {};
        
    \draw[rot, thick] (S5) -- ($(S5) + 0.2*(SNmin1) - 0.2*(S5)$);
    \draw[rot, thick] (SNmin1) -- ($(S5) + 0.8*(SNmin1) - 0.8*(S5)$);
        
    \path (S5) -- node[rot, sloped]{\large\ldots} (SNmin1);
        
    \draw[rot, thick] (start) -- (S1) -- (S2) -- (S3) -- (S4) -- (S5);
        
    \draw[rot, thick] (SNmin1) -- (end);
        
    \draw[->] (0,0) -- (0,2.5);
    \draw[->] (0,0) -- (6,0);
    \node at (-.25,2.25) {\footnotesize$\mu$};
    \foreach \i in {1,...,6}
    {
        \draw ($(0.5*\i, 0.08)$) -- ($(0.5*\i, -0.08)$);
        \pgfmathtruncatemacro{\x}{(\i - 1)};
        \node at ($(0.5*\i, -0.25)$) {\footnotesize $\x$};
    }
    \draw (5.0, 0.08) -- (5.0, -0.08);
    \draw (5.5, 0.08) -- (5.5, -0.08);
    \node at (5.5,-0.25) {\footnotesize $N$};
    \node at (5.0,-0.25) {\footnotesize $N\kern-.3em-\kern-.3em1$};
    \node at (6.25,0) {\footnotesize$K$};
        
    \node [orange] (mu) at (4,0.4) {\footnotesize$\mu(\mathcal{S})$};
        
    \node [rot, inner sep = 0.5pt, outer sep=1pt] (S) at (0.6,1.2) {\footnotesize$\mathcal{S}$};
    \draw [rot] (S) -- ($0.5*(start) + 0.5*(S1)$);
        
\end{tikzpicture}}
    \caption{\footnotesize Graph of polar code designs with performance metric $\mu$. The optimal sequence $\mathcal{S}$ minimizes the area under the graph.}
    \label{fig:graph}
    \vspace{-.5cm}
\end{figure}

\subsection{Rate Compatible Symmetric Polar Codes}
For a given symmetry constraint, single bit-granular sequences of polar codes do not exist \cite{Geiselhart2023ratecompatible}. Even when dropping hard symmetry constraints, it is unlikely that a single bit-granular sequence with satisfactory performance under \ac{AED} can be constructed.
For versatile communication systems, however, such flexibility in the code dimension is desired.
Fortunately, the coding gain achieved by certain polar codes under \ac{AED} is so large that zero-padding for minor rate-adaption becomes viable.
That is, if no code with the desired dimension $K$ fulfills the requirements, find the next larger allowed $K'$ and append $z=K'-K$ zeros to each message.

\begin{definition}[Zero-Padded Polar Code]\label{def:zeropadding}\normalfont
Let $\mathcal{P}(\mathcal{I},z)$ denote the code with dimension $K = |\mathcal{I}|-z$ that is obtained by padding the messages $\boldsymbol{u}\in \FF_2^K$ with $z$ zeros and encoding it using $\mathcal{P}(\mathcal{I}$).
\end{definition}
\begin{proposition}[Performance Bound for Zero-Padded Polar Codes]\label{prop:zpperformance}\normalfont
Let $\mu(\mathcal{C})$ denote the required $E_\mathrm{s}/N_0$ to decode a code $\mathcal{C}$ with block-error rate $\epsilon$. 
Then $\mu(\mathcal{P}(\mathcal{I},z)) \le \mu(\mathcal{P}(\mathcal{I})) $.
\end{proposition}
\begin{IEEEproof}
By definition, $\mathcal{P}(\mathcal{I},z)$ is a subcode of $\mathcal{P}(\mathcal{I})$. Thus, one can decode  $\mathcal{P}(\mathcal{I},z)$ with any decoder for $\mathcal{P}(\mathcal{I})$, since the decoder does not need to 
be aware of the fact 
that only a subset of messages (those with zeros in the zero-padded positions) is transmitted. Hence, at the same $E_\mathrm{s}/N_0$, the same \ac{BLER} is obtained. Decoders utilizing the zero-padded bits will only improve on this performance.
\end{IEEEproof}

While in theory, the position of the zeros does not matter, for the purpose of rate-compatible sequences the zero-padding should be in unreliable positions, so that the smaller codes are subcodes of the larger codes. 
In the following remarks, we give ideas how additional gains are generated from these bits.
\begin{remark}\label{rk:betterdecoding}\normalfont
    In its simplest form, the decoder decodes the larger code with dimension $K'$ without ``knowing'' about the zero-padding. Utilizing the zero-padded bits as frozen bits (in the non-permuted decoding path) will further improve the performance.
\end{remark}
\begin{remark}\label{rk:systematic}\normalfont
    When systematic encoding is used, the zero-padded bits will show up as zeros in the codeword, and thus, all decoding paths may use this as prior information, i.e., \ac{LLR} values of $+\infty$.
\end{remark}
\begin{remark}\normalfont
    Instead of zero-padding, also an error-detecting outer code (such as parity bits, \ac{CRC}, etc.) may be employed to improve the false alarm rate.
\end{remark}

Based on this idea, we augment the graph by adding vertices corresponding to zero-padded polar codes up to a sufficiently large maximum zero-padding length $z_\mathrm{max}$. 
The connectivity rule is extended as follows:
\begin{definition}[Augmented Graph]\normalfont\label{def:graph}
Two vertices $\mathcal{P}(\mathcal{I}_i,z_i)$ and $\mathcal{P}(\mathcal{I}_j,z_j)$ are connected by an edge, if and only if
\begin{itemize}
    \item $|\mathcal{I}_j|-z_j = |\mathcal{I}_i|-z_i +1$
    \item $\mathcal{I}_i = \mathcal{I}_j$ or ($z_i=0$ and $\mathcal{I}_i \subseteq \mathcal{I}_j$).
\end{itemize}
\end{definition}

The first condition ensures the single-bit granularity, while the second is the subcode property.
In Fig.~\ref{fig:graph}, the purple vertex corresponds to zero-padding the best code with $K=4$, creating a ``shortcut'' for better overall performance metric. It is located higher than $\mathcal{P}(\mathcal{I}_4)$, as $\mu$ shall represent the $E_\mathrm{b}/N_0$ value in dB, which exhibits a degradation due to the rate-loss induced by the zero-padding. From the worst-case bound given Proposition~\ref{prop:zpperformance}, this logarithmic performance measure in dB can be expressed as
\begin{equation}\label{eq:rateloss}
    \mu(\mathcal{P}(\mathcal{I},z)) = \mu(\mathcal{P}(\mathcal{I})) + 10\log_{10} \frac{|\mathcal{I}|}{|\mathcal{I}|-z}.
\end{equation}
Hence, no additional Monte-Carlo simulations are required.
Note that the position of the zero-padded bits is left open and will be automatically determined (up to permutation) by the search algorithm.
However, due to the ambiguity in the order of the zero-padded bits, only a partial order $P$ is obtained.
The search algorithm with zero padding is given in Algorithm~\ref{alg:rate_comp_construction}.

\begin{algorithm}[tbh]
	\SetAlgoLined
	\LinesNumbered
	\SetKwInOut{Input}{Input}\SetKwInOut{Output}{Output}
	\Input{Block length $N$, dataset of polar codes $\mathcal{D}$ with performance metric $\mu$, $z_\mathrm{max}$.}
	\Output{Partial order $P$.}
	Augment dataset by $z\in \{0,...,z_\mathrm{max}\}$\;
	Construct graph according to Def.~\ref{def:graph}\;
	Assign edge costs according to (\ref{eq:edgecost})\;
	Compute shortest path\;
	Extract partial order $P$\;
\caption{\footnotesize Construction of the optimal rate-compatible sequence.\label{alg:rate_comp_construction}}
\end{algorithm}
\vspace{-.5cm}

\section{Nested Sequences}
To incorporate multiple block lengths $N$, the above procedure is extended in a sequential manner.
While in general, a global search would be best, the complexity does not scale well due to the zero padding.
For each code, all contained shorter subcodes must be considered, each either directly or as a zero-padded larger code.
This is typically computationally infeasible.
However, several, sub-optimal meta-search strategies are possible, e.g., a sequential approach, as outlined in the following.
\vspace{-.2cm}
\subsection{Sequential Construction}
We propose to sequentially optimize each block length $N$, incorporating the partial orders found in the previous steps.
To this end, the dataset is filtered to only include codes compliant with the constraints posed by the partial order $P$. Then, Algorithm \ref{alg:rate_comp_construction} is used to find the optimal ordering for $N$, and $P$ is updated to include the newly found relations.
This procedure is given in Algorithm~\ref{alg:nested_construction}.

\begin{algorithm}[tbh]
	\SetAlgoLined
	\LinesNumbered
	\SetKwInOut{Input}{Input}\SetKwInOut{Output}{Output}
	\Input{Block length schedule $\boldsymbol{N}$, dataset of polar codes $\mathcal{D}$ with performance metric $\mu$, $z_\mathrm{max}$.}
	\Output{Partial order $P$.}
	$P \gets \emptyset$\;
	\For {$N \in \boldsymbol{N}$}{
    	Filter dataset according to $P$\;
    	$P' \gets$ Algorithm \ref{alg:rate_comp_construction}($N$,\,$\mathcal{D}$,\,$z_\mathrm{max}$)\;
    	$P \gets P \cup P'$\;
	}
\caption{\footnotesize Construction of nested rate-compatible sequence.\label{alg:nested_construction}}
\end{algorithm}
\vspace{-.4cm}

\subsection{Disambiguation}
Due to the zero-padding, the partial order obtained from Algorithm~\ref{alg:nested_construction} is not necessarily a total order, i.e., it contains ambiguities. 
For decoding with decoders unaware of the zero-padding, these ambiguities are irrelevant.
However, there are several reasons to specify a total order:
\begin{itemize}
    \item Unique specification of the encoding.
    \item Improved performance when the decoding strategy outlined in Remark~\ref{rk:betterdecoding} is applied.
    \item Applicability of other decoding methods other than the one the code is designed for, e.g., \ac{SCL}.
\end{itemize}
Therefore, we now outline a methodology to construct a total order, using auxiliary bit channel reliabilities $\boldsymbol{\nu}$ that incorporate the above listed factors.
Namely, the partial order is traversed breadth-first, and whenever there is an ambiguity, the auxiliary channel reliabilities $\boldsymbol{\nu}$ are used to determine a total ordering.
Algorithm~\ref{alg:total_order} lists this procedure in pseudocode.
For example, the auxiliary channel reliabilities $\boldsymbol{\nu}$ may be obtained from density evolution \cite{DensityEvolution}.
Also, in the same way, longer sequences may be obtained that optimize for other metrics, such as performance under different decoding algorithms.
\vspace{-.2cm}
\begin{algorithm}[tbh]
	\SetAlgoLined
	\LinesNumbered
	\SetKwInOut{Input}{Input}\SetKwInOut{Output}{Output}
	\Input{Partial order $P$, bit channel reliabilities $\boldsymbol{\nu}$.}
	\Output{Total order $Q$.}
	\For {$i=0,\dots,N-1$}{
	    $\mathcal{T} \gets \text{ possible successors of } Q_{0:i-1} \text{ based on } P$\;
	    $q_i \gets \arg \max_{j\in \mathcal{T}} \nu_j $\;
	}
\caption{\footnotesize Disambiguation.\label{alg:total_order}}
\end{algorithm}
\vspace{-.5cm}

\section{Numerical Results}
In this section, we demonstrate the proposed design methodology for the case of automorphism ensemble (AE) \ac{SC} decoding. 
As this decoding algorithm is most effective for short and medium length polar codes \cite{Geiselhart2023ratecompatible}, we consider block lengths $32\le N \le256$.
All results are obtained for the \ac{AWGN} channel with \ac{BPSK} modulation, and the target \ac{BLER} is set to $\epsilon=10^{-3}$.
The ensemble size is $M=8$ to be comparable to the standard list size for \ac{SCL} decoding, for which the 5G polar code has been designed \cite{polarDesign5G}, and the permutations are randomly selected from the \ac{BLTA} group for each code.

\subsection{Datasets}
For block lengths $N\in \{32, 64, 128$\}, it is feasible to enumerate all polar codes following the \ac{UPO} (118, 1172 and 25728 codes, respectively) and evaluate their performance under AE-SC-8 decoding using Monte-Carlo simulation. 
For $N=256$, the dataset consists of all 11897 codes with up to two elements in $\mathcal{I}_\mathrm{min}$, as they contribute the most symmetric polar codes \cite{polar_aed}. 
Additionally, referring to Proposition~\ref{prop:supercode}, codes constructed from the information sets of the best shorter codes are included in the dataset.
All datasets record the required $E_\mathrm{b}/N_0$ in dB required to achieve the target \ac{BLER} as the performance metric $\mu$ for each of the codes.

\subsection{Error-Rate Performance}
Using Algorithm~\ref{alg:nested_construction}, we design nested polar codes both with increasing schedule ($\boldsymbol{N}=[32,64,128,256]$) and decreasing schedule ($\boldsymbol{N}=[256,128,64,32]$), respectively.
As a reference, also a non-nested (unconstrained) design is evaluated, which applies Algorithm~\ref{alg:rate_comp_construction} for each value of $N$ individually.
In each case, the dataset is augmented using zero-padding with sufficiently large $z_\mathrm{max}$ according to Definition~\ref{def:zeropadding}.
Each zero-padded code is assigned the $\mu$-value of its containing supercode, adjusted for the rate-loss induced by the zero-padding according to Equation (\ref{eq:rateloss}).

\begin{table}[htb]
    \caption{\footnotesize Average $E_\mathrm{b}/N_0$ in dB to achieve BLER $ = 10^{-3}$ for all code dimensions $K$ of different rate compatible polar code designs. %
    }
    \centering
    \begin{NiceTabular}{c|c|cccc}
        \multirow{2}[2]{*}{$N$} & CA-SCL-8 & \multicolumn{4}{c}{AE-SC-8} \\ 
        \cmidrule(lr){2-2}\cmidrule(lr){3-6}
        & 5G & ZP-inc & ZP-dec & unconstr. & $\beta_{\boldsymbol{s}}$ \cite{Geiselhart2023ratecompatible}  \\
        \midrule
         32 &         5.7770  & \textbf{5.7403} & 5.8194 & 5.7403 & 5.7972 \\
         64 &         5.7114  & \textbf{5.1055} & 5.2377 & 5.1035 & 5.2030 \\
        128 &         4.6771  & \textbf{4.4924} & 4.5547 & 4.4859 & 4.6052 \\
        256 & \textbf{3.7992} &         4.0180  & 3.9539 & 3.9539 & 4.0519 \\
        \midrule
        $\sum$ & 19.9647 & \textbf{19.3563} & 19.5657 & 19.2836 & 19.6572 \\
    \end{NiceTabular}
    \vspace{-0.2cm}
    \begin{equation*}
        \mspace{92.25mu}\underbrace{\hphantom{19.9647||||||19.5657||||||19.3816}}_{\text{\scriptsize Nested}}
        \mspace{16.5mu}\underbrace{\hphantom{19.9647||||||19.5657}}_{\text{\scriptsize Non-Nested}}\mspace{100mu}
    \end{equation*}
    \label{tab:metric}
    \vspace{-.8cm}
\end{table}

Table~\ref{tab:metric} lists the average value of $\mu$ (i.e., $\mu(\mathcal{S})/N$) obtained for each constructed sequence for each block length.
Their sum is also listed, as a heuristic for comparing the different design procedures.
Finally, we also give these values for the 5G polar code under \ac{CRC}-aided \ac{SCL} decoding and the symmetric $\beta$-expansion proposed in \cite{Geiselhart2023ratecompatible} under AE-SC-8 decoding.
Note that for the latter, zero-padding has also been applied to obtain a 1-bit granular design.
As we can see, for all block lengths except $N=256$, the incrementing schedule of the proposed nested code design algorithm gives the best performance, very close to the individually optimized, non-nested designs. It also outperforms the symmetric $\beta$-expansion design for all lengths.

Fig.~\ref{fig:perf} shows the required \ac{SNR} to achieve the target \ac{BLER} $\epsilon=10^{-3}$ for the proposed nested polar code design under AE-SC-8 decoding and compares it to the 5G polar code under \ac{CA-SCL} decoding, for different values of $N$ and $K$.
The plot also shows the $\mathcal{O}(n^{-2})$ approximation of the finite block length \ac{PPV} meta-converse bound \cite{ErsegheMetaConverse} as a baseline.
The proposed design has a very similar performance as the 5G polar code, with notable performance gains in for small code dimensions, all while using the much more efficient AE-\ac{SC} decoding algorithm.
Note the kinks in the curves at the code dimensions of first-order \ac{RM} codes, which are well-suited for AE-\ac{SC} decoding and are not obtained by the \ac{CRC}-aided 5G construction.
The small plateaus in the curves are due to the zero padding, i.e., the performance of the code is identical to its containing supercode in terms of $E_\mathrm{s}/N_0$.
For longer lengths, \ac{AED} becomes less effective, while the rate-loss of the \ac{CRC} decreases, resulting in better performance of \ac{CA-SCL} decoding.

\begin{figure}[htb]
    \centering
    \resizebox{\linewidth}{!}{\input{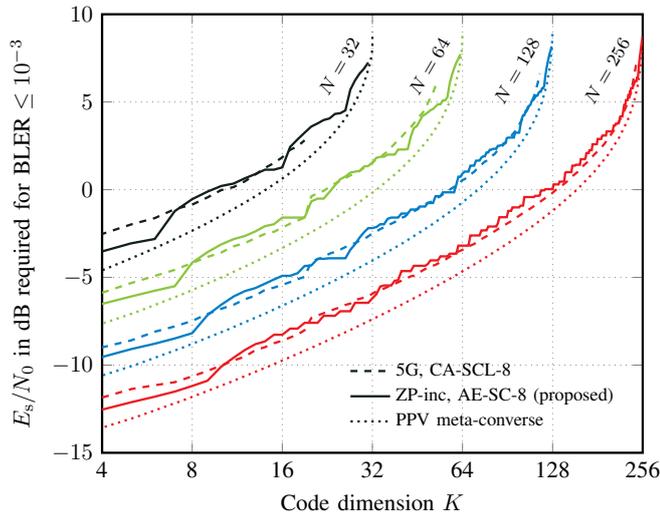}}
    \caption{\footnotesize Required SNR for BLER $ \le 10^{-3}$ vs. code dimension $K$ for different block lengths.}
    \label{fig:perf}
\end{figure}

\subsection{Disambiguated Total Order}
The designed nested sequence (using the incrementing schedule) is converted to a total order $Q$ using Algorithm~\ref{alg:total_order}, with auxiliary bit channel reliabilities obtained from density evolution.
This total order is listed in Table~\ref{tab:Q} and is read row-wise from left to right.
\begin{table}[htb]
    \caption{\footnotesize Total order $Q$ of the nested rate-compatible polar code optimized for AE-SC-8 decoding.}
    \centering
    \setlength{\tabcolsep}{2pt}

\resizebox{\linewidth}{!}{
\footnotesize\begin{tabular}{rrrrrrrrrrrrrrrr}
0 & 1 & 2 & 4 & 8 & 16 & 32 & 64 & 128 & 3 & 5 & 6 & 9 & 17 & 33 & 65\\
129 & 10 & 12 & 18 & 34 & 66 & 130 & 20 & 24 & 36 & 40 & 48 & 68 & 72 & 80 & 96\\
132 & 136 & 144 & 160 & 192 & 7 & 11 & 19 & 35 & 67 & 131 & 13 & 14 & 21 & 37 & 69\\
133 & 22 & 38 & 70 & 134 & 25 & 41 & 49 & 73 & 81 & 137 & 97 & 145 & 161 & 193 & 26\\
28 & 42 & 50 & 44 & 52 & 56 & 74 & 82 & 138 & 98 & 146 & 162 & 194 & 76 & 84 & 100\\
88 & 104 & 112 & 15 & 23 & 39 & 71 & 135 & 140 & 148 & 164 & 196 & 152 & 168 & 176 & 200\\
208 & 224 & 27 & 43 & 51 & 75 & 83 & 99 & 139 & 147 & 163 & 195 & 29 & 30 & 45 & 46\\
77 & 78 & 141 & 142 & 53 & 85 & 101 & 149 & 165 & 197 & 54 & 86 & 102 & 150 & 166 & 198\\
57 & 89 & 105 & 153 & 169 & 201 & 113 & 177 & 209 & 225 & 58 & 90 & 106 & 154 & 170 & 202\\
114 & 178 & 210 & 226 & 60 & 92 & 108 & 116 & 120 & 31 & 47 & 79 & 55 & 87 & 103 & 156\\
172 & 180 & 204 & 212 & 228 & 143 & 151 & 167 & 199 & 184 & 216 & 232 & 240 & 59 & 91 & 107\\
155 & 171 & 203 & 115 & 179 & 211 & 227 & 61 & 62 & 93 & 94 & 157 & 158 & 109 & 110 & 173\\
174 & 205 & 206 & 117 & 181 & 213 & 229 & 118 & 182 & 214 & 230 & 121 & 185 & 217 & 233 & 241\\
122 & 124 & 63 & 186 & 188 & 95 & 159 & 218 & 220 & 234 & 242 & 111 & 175 & 207 & 236 & 244\\
248 & 119 & 183 & 215 & 231 & 123 & 125 & 126 & 187 & 189 & 190 & 219 & 221 & 222 & 235 & 243\\
237 & 238 & 245 & 249 & 246 & 250 & 252 & 127 & 191 & 223 & 239 & 247 & 251 & 253 & 254 & 255
\end{tabular}
}
    \label{tab:Q}
    \vspace{-.5cm}
\end{table}

\section{Conclusion and Outlook}
In this paper, we derived theoretical properties of the nesting structure of polar codes with symmetries and proposed a data-driven methodology for constructing rate-compatible polar codes with and without a nesting property. 
Although the proposed method is general, it is well suited for symmetric polar codes, as the symmetry constrains the search space to make a fully data-driven approach feasible.
The algorithms are demonstrated by constructing the first nested family of rate-compatible polar codes for \ac{AED}.
Using Monte-Carlo simulations, we show that the designed polar codes outperform existing rate-compatible code designs.
The proposed design is also compatible with \ac{CA-SCL} and also \ac{SCAL} \cite{Johannsen2023SCAL} decoding.
Future work includes the fine adjusting of the block length by shortening or puncturing, e.g., using the same length matching procedure standardized in 5G
and applying \ac{AED} based on the mother code.
Also, the approach using the automorphism groups of the shortened codes proposed in \cite{Pillet2024Shortened} may be worth exploring in the context of nested polar codes.
Furthermore, as the proposed design methodology is very general, also other constraints, such as the set of optimized leaf node decoders for fast simplified \ac{SC} decoding can be incorporated.

\bibliographystyle{IEEEtran}
\bibliography{references.bib}

\begin{thebibliography}{10}
\providecommand{\url}[1]{#1}
\csname url@samestyle\endcsname
\providecommand{\newblock}{\relax}
\providecommand{\bibinfo}[2]{#2}
\providecommand{\BIBentrySTDinterwordspacing}{\spaceskip=0pt\relax}
\providecommand{\BIBentryALTinterwordstretchfactor}{4}
\providecommand{\BIBentryALTinterwordspacing}{\spaceskip=\fontdimen2\font plus
\BIBentryALTinterwordstretchfactor\fontdimen3\font minus
  \fontdimen4\font\relax}
\providecommand{\BIBforeignlanguage}[2]{{%
\expandafter\ifx\csname l@#1\endcsname\relax
\typeout{** WARNING: IEEEtran.bst: No hyphenation pattern has been}%
\typeout{** loaded for the language `#1'. Using the pattern for}%
\typeout{** the default language instead.}%
\else
\language=\csname l@#1\endcsname
\fi
#2}}
\providecommand{\BIBdecl}{\relax}
\BIBdecl

\bibitem{ArikanMain}
E.~{Arıkan}, ``{Channel Polarization: A Method for Constructing
  Capacity-Achieving Codes for Symmetric Binary-Input Memoryless Channels},''
  \emph{IEEE Trans. Inf. Theory}, vol.~55, no.~7, pp. 3051--3073, Jul. 2009.

\bibitem{polar5G2018}
{Technical Specification Group Radio Access Network}, ``{3GPP, 2018, TS 38.212
  V.15.1.1.}''

\bibitem{rowshan2024channelcoding6g}
M.~Rowshan, M.~Qiu, Y.~Xie, X.~Gu, and J.~Yuan, ``{Channel Coding Toward 6G:
  Technical Overview and Outlook},'' \emph{IEEE Open Journal of the
  Communications Society}, vol.~5, pp. 2585--2685, 2024.

\bibitem{proceedings2024trends}
S.~Miao, C.~Kestel, L.~Johannsen, M.~Geiselhart, L.~Schmalen,
  A.~Balatsoukas-Stimming, G.~Liva, N.~Wehn, and S.~ten Brink, ``{Trends in
  Channel Coding for 6G},'' \emph{Proceedings of the IEEE}, pp. 1--23, 2024.

\bibitem{rm_automorphism_ensemble_decoding}
M.~Geiselhart, A.~Elkelesh, M.~Ebada, S.~Cammerer, and S.~ten Brink,
  ``{Automorphism Ensemble Decoding of Reed--Muller Codes},'' \emph{IEEE Trans.
  Commun.}, vol.~69, no.~10, pp. 6424--6438, 2021.

\bibitem{pilletPolarCodesForAED}
C.~Pillet, V.~Bioglio, and I.~Land, ``{Polar Codes for Automorphism Ensemble
  Decoding},'' in \emph{IEEE Inf. Theory Workshop (ITW)}, 2021.

\bibitem{polar_aed}
M.~Geiselhart, A.~Elkelesh, M.~Ebada, S.~Cammerer, and S.~ten Brink, ``{On the
  Automorphism Group of Polar Codes},'' in \emph{IEEE Inter. Symp. Inf. Theory
  (ISIT)}, 2021, pp. 1230--1235.

\bibitem{Kestel2023URLLC}
C.~Kestel, M.~Geiselhart, L.~Johannsen, S.~ten Brink, and N.~Wehn,
  ``{Automorphism Ensemble Polar Code Decoders for 6G URLLC},'' in \emph{13th
  ITG Inter. Conf. on Systems, Comm., and Coding (SCC)}, 2023.

\bibitem{pillet2023distribution}
C.~Pillet and V.~Bioglio, ``{On the Distribution of Partially-Symmetric Codes
  for Automorphism Ensemble Decoding},'' in \emph{2023 IEEE Information Theory
  Workshop (ITW)}, 2023, pp. 36--41.

\bibitem{bits2023unified}
M.~Geiselhart, F.~Krieg, J.~Clausius, D.~Tandler, and S.~ten Brink, ``{6G: A
  Welcome Chance to Unify Channel Coding?}'' \emph{IEEE BITS the Information
  Theory Magazine}, vol.~3, no.~1, pp. 67--80, 2023.

\bibitem{Korada2010optimal}
S.~B. Korada and R.~L. Urbanke, ``Polar codes are optimal for lossy source
  coding,'' \emph{IEEE Transactions on Information Theory}, vol.~56, no.~4, pp.
  1751--1768, 2010.

\bibitem{polarDesign5G}
V.~Bioglio, C.~Condo, and I.~Land, ``{Design of Polar Codes in 5G New Radio},''
  \emph{ArXiv e-prints}, Apr. 2018.

\bibitem{Huang2019NestedReinforcement}
L.~Huang, H.~Zhang, R.~Li, Y.~Ge, and J.~Wang, ``{Reinforcement Learning for
  Nested Polar Code Construction},'' in \emph{2019 IEEE Global Communications
  Conference (GLOBECOM)}, 2019, pp. 1--6.

\bibitem{Li2021LearningNested}
Y.~Li, Z.~Chen, G.~Liu, Y.-C. Wu, and K.-K. Wong, ``{Learning to Construct
  Nested Polar Codes: An Attention-Based Set-to-Element Model},'' \emph{IEEE
  Communications Letters}, vol.~25, no.~12, pp. 3898--3902, 2021.

\bibitem{ankireddy2024nestedtransformer}
S.~K. Ankireddy, S.~A. Hebbar, H.~Wan, J.~Cho, and C.~Zhang, ``{Nested
  Construction of Polar Codes via Transformers},'' in \emph{2024 IEEE Int.
  Symposium on Information Theory (ISIT)}, 2024, pp. 1409--1414.

\bibitem{pillet2022AEDdesign}
C.~Pillet, V.~Bioglio, and I.~Land, ``{Classification of Automorphisms for the
  Decoding of Polar Codes},'' in \emph{IEEE Inter. Conf. on Commun. (ICC)},
  2022, pp. 110--115.

\bibitem{ShabunovPredefinedAutomorphisms}
K.~Shabunov, ``{Monomial Codes With Predefined Automorphisms},'' in \emph{2022
  IEEE/CIC International Conference on Communications in China (ICCC
  Workshops)}, 2022, pp. 205--210.

\bibitem{Geiselhart2023ratecompatible}
M.~Geiselhart, J.~Clausius, and S.~ten Brink, ``{Rate-Compatible Polar Codes
  for Automorphism Ensemble Decoding},'' in \emph{2023 12th International
  Symposium on Topics in Coding (ISTC)}, 2023, pp. 1--5.

\bibitem{geiselhart2022graphsearch}
M.~Geiselhart, A.~Zunker, A.~Elkelesh, J.~Clausius, and S.~ten Brink, ``{Graph
  Search based Polar Code Design},'' in \emph{Asilomar Conference on Signals,
  Systems and Computers}, Nov. 2022.

\bibitem{bardet_polar_automorphism}
M.~{Bardet}, V.~{Dragoi}, A.~{Otmani}, and J.~{Tillich}, ``{Algebraic
  Properties of Polar Codes From a New Polynomial Formalism},'' in \emph{IEEE
  Inter. Symp. Inf. Theory (ISIT)}, 2016, pp. 230--234.

\bibitem{LiBLTA2021}
Y.~Li, H.~Zhang, R.~Li, J.~Wang, W.~Tong, G.~Yan, and Z.~Ma, ``The complete
  affine automorphism group of polar codes,'' in \emph{IEEE Global Commun.
  Conf. (GLOBECOM)}, Dec. 2021.

\bibitem{talvardyList}
I.~Tal and A.~Vardy, ``{List Decoding of Polar Codes},'' \emph{IEEE Trans. Inf.
  Theory}, vol.~61, no.~5, pp. 2213--2226, May 2015.

\bibitem{moriDE}
R.~Mori and T.~Tanaka, ``{Performance and Construction of Polar Codes on
  Symmetric Binary-Input Memoryless Channels},'' in \emph{2009 IEEE
  International Symposium on Information Theory}, 2009, pp. 1496--1500.

\bibitem{Dijkstra1959}
E.~W. Dijkstra, ``A note on two problems in connexion with graphs,''
  \emph{Numerische Mathematik}, vol.~1, no.~1, pp. 269--271, 1959.

\bibitem{DensityEvolution}
R.~Mori and T.~Tanaka, ``{Performance of Polar Codes with the Construction
  using Density Evolution},'' \emph{IEEE Communications Letters}, vol.~13,
  no.~7, pp. 519--521, 2009.

\bibitem{ErsegheMetaConverse}
T.~Erseghe, ``{Coding in the Finite-Blocklength Regime: Bounds Based on Laplace
  Integrals and Their Asymptotic Approximations},'' \emph{IEEE Trans. Inf.
  Theory}, vol.~62, no.~12, pp. 6854--6883, 2016.

\bibitem{Johannsen2023SCAL}
L.~Johannsen, C.~Kestel, M.~Geiselhart, T.~Vogt, S.~ten Brink, and N.~Wehn,
  ``{Successive Cancellation Automorphism List Decoding of Polar Codes},'' in
  \emph{2023 12th International Symposium on Topics in Coding (ISTC)}, 2023,
  pp. 1--5.

\bibitem{Pillet2024Shortened}
C.~Pillet, I.~Sagitov, V.~Bioglio, and P.~Giard, ``{Shortened Polar Codes Under
  Automorphism Ensemble Decoding},'' \emph{IEEE Communications Letters},
  vol.~28, no.~4, pp. 773--777, 2024.

\end{thebibliography}
\end{NoHyper}
\end{document}